# Why it has become more difficult to predict Nobel Prize winners: a bibliometric analysis of Nominees and Winners of the Chemistry and Physics Prizes (1901-2007)

Yves Gingras and Matthew L. Wallace

Observatoire des sciences et des technologies (OST), Centre interuniversitaire de recherche sur la science et la technologie (CIRST), Université du Québec à Montréal, Case Postale 8888, succ. Centre-Ville, Montréal (Québec), H3C 3P8, Canada. E-mail: gingras.yves@uqam.ca; mattyliam@gmail.com

**Abstract**

We propose a comprehensive bibliometric study of the profile of Nobel prizewinners in chemistry and physics from 1901 to 2007, based on citation data available over the same period. The data allows us to observe the evolution of the profiles of winners in the years leading up to – and following – nominations and awarding of the Nobel Prize. The degree centrality and citation rankings in these fields confirm that the Prize is awarded at the peak of the winners' careers, despite brief a Halo Effect observable in the years following the attribution of the Prize. Changes in the size and organization of the two fields result in a rapid decline of predictive power of bibliometric data over the century. This can be explained not only by the growing size and fragmentation of the two disciplines, but also, at least in the case of physics, by an implicit hierarchy in the most legitimate topics within the discipline, as well as among the scientists selected for the Prize. Furthermore, the lack of readily-identifiable dominant contemporary physicists suggests that there are few new paradigm shifts within the field, as perceived by the scientific community as a whole.

**Introduction**

In science, the prestige and status associated with the Nobel Prize is unmatched. An important symbol of scientific achievement and discovery, the Prize annually generates enormous interest in the scientific community and the general public. Accordingly, many studies have been devoted to the institutional aspects of the history of the Prize [Zuckerman 1977; Crawford 1984; Friedman, 2001] as well as to the historical study of particular winners, [Barkan 1994; Björk 2001; Jenkins 2001; Krige 2001; Elzinga 2006]. Also, thanks to the opening of the Nobel archives for prizes and nominees dating back more than 50 years, we better know the population of scientists nominated for the coveted prizes [Bernhard et al, 1982; MacLachlan, 1991; Crawford 1992].

From the point of view of bibliometrics, one could ask if Nobelists are more cited than the average scientist and if we can find a particular pattern for winners that would distinguish them from the rest of the community. In the latter case, one could even try to "guess", or predict, the next winner. Eugene Garfield has explored some of these questions in a series of papers attempting to elucidate the profile of prizewinners by describing a subset of scientists "of Nobel class" via their citation statistics [Garfield 1977, 1986; Garfield & Welljams-Dorof, 1992]. Not surprisingly, it was found that this set of scientists does differ in citation frequency from the "average" scientist: "in the highest percentile [...] a significant percentage have won the Nobel Prize or go on to win the Prize in later years. Also, the author impact of Nobelists is sufficiently high to distinguish them from non-Nobelists" [Garfield & Welljams-Dorof, 1992, p.118]. Garfield also claims that Nobel laureates can be distinguished from other scientists by having written "citation classics". It should be noted, however, that while Garfield claims that bibliometrics have substantial predictive power, he admits that the subjectivity of the Nobel selection process by definition precludes any *systematic* forecasting from "objective" data [Garfield, 1986]. A similar paper by Ashton and Oppenheim [1978] presents an improvement on Garfield's method by, among other things, generalizing the citation statistics to include non-first authors. They also claim not to be able to predict a prizewinner, but rather to identify a group of candidates likely to win the Prize. In general, the inclusion of multi-authored papers improves the rankings, but does not substantially alter them. It should also be noted that most of the predictions discussed above were done *a posteriori*. In the same vein, Kademani et al. [2005] have demonstrated the dominance of multi-authored papers by Nobel laureates, although they use data from only 8 winners. Other similar bibliometric studies have also restricted their scope to a few selected laureates [Kademani, 2005 and references therein]. Finally, a recent article by Karazua and Momkausaité [2004] has brought to light certain characteristics of the Nobel Prize in physics, most notably the distribution of winners' ages, their fields of study and the lag times between a discovery and its corresponding Prize.

An important aspect that is missing from most bibliometric analysis of Nobel prizewinners is a sensitivity to the evolution *in time* of the dynamics and growth of science that may affect the pattern of citations and the relative position of prizewinners in the structure of the scientific field. It is obvious that science has grown exponentially over the $20^{th}$ century and that all disciplines have given rise to many specialties to such an extent that the fragmentation of science makes it more difficult now than ever to identify an obvious winner for a discipline as a whole. Whereas it was still relatively easy around 1910 to know who the most important scientists in a discipline were, such a judgment is much more difficult since at least the 1970s. In order to analyze the effect of these changes on the distribution of citation to prospective winners of a Nobel Prize, we provide in this paper a detailed analysis of the changing pattern of citation and centrality in the co-citation network of all winners of the chemistry and physics prizes from 1901 to 2007, as well as for nominees (1901-1945). As we will see, the changing dynamic of science and its obvious growth and increased specialization since the 1960s had the effect of diluting potential winners in a massive group of central scientists, from which it has become nearly impossible to pick a winner using bibliometric tools.

**Methodology**

The basis for this paper is a list of the 500 most cited and most central[*] chemists and physicists constructed annually from the citation data of the subset of physics and chemistry journals in the Web of Science for the period 1900-2006.[†] We thus exclude important multidisciplinary journals such as *Nature* and *Science*, but ensure a good representation of the disciplinary journals for each field. The two measures of a scientists' importance are distinct although highly correlated: the more citations an author receives, the more chance he has of becoming "central" in the network of co-citations. It terms of co-citations, one can interpret centrality as a measure of an author's position in the discipline's network [Gingras, 2007]. Using rankings instead of absolute numbers allows us to have a time invariant measure of the most influential scientists in a given year. We then compare this data with the list of all Nobel prizewinners between 1901 and 2007, as well as with the nominees for the period 1901-1945 [Crawford 2002].

---

[*] Centrality is defined as Freeman's degree centrality of author in a co-citation network. See [Freeman, 1977/1978].

[†] Although we have analyzed the data for the Nobel prize in medicine, we chose not to include them in this paper. Our hypothesis is that the selection of winners is performed in a very distinct manner from chemistry or physics, and often depends on practical applications instead of published "discoveries" as such. The citation patterns and statistics are therefore very different in medicine than in physics and chemistry.

From a bibliometric point of view, one would expect that Nobel Prizes are awarded to the most cited authors, or are at least chosen among the most cited, since it is taken as nearly axiomatic that citations constitute a good indicator of the recognition received by scientists and thus of their global symbolic capital in the scientific field [Cole & Cole, 1973; Bourdieu, 1975]. Using our large data sample composed of 330 winners in physics and chemistry (1901-2007) and 1595 nominees (including "repeat" nominees) in the same two disciplines (1901-1945) and comparing their results with the 500 most cited scientists in their field, we can examine in greater detail the profile of prizewinners and observe the changing characteristics of the distribution of ranks over a large time period before as well as after their being nominated or having received the Prize. For example, for each prizewinner, we are able to obtain his ranking in terms of citation and centrality the year his Prize was awarded, as well as in the years preceding and following the event. This yields a large time interval for each author, and we can average the results over all prizewinners (or nominees), setting the year "0" as the winning (or nomination) year for each one. When someone wins twice (a rare event) or is nominated many times (a more frequent event), we can use the same citation data many times, each time with respect to the year of the Prize or nomination in question. In all cases, care must be taken to make sure the data contain minimal namesakes, not only in terms of the laureates themselves, but all prominent scientists as well, in order to ensure that the rankings are accurate. Often, in order to display the data in a meaningful manner, the rankings (from 1 to 500) are inverted. We have applied a "cutoff" at 500, beyond which it becomes very computationally time-consuming to collect data. However, given the usual distribution of citations, neglecting scientists who rank below 500 does not affect the results. In order to check this point, we have performed a similar analysis using 100 as a cutoff and, as expected, we have obtained statistically similar results.

**Nominees, laureates and the development of disciplines: 1901-1945**

Using the method discussed above, Figures 1 and 2 display the average rank in terms of citations and centrality of Nobel laureates and nominees in chemistry and physics for the period 1901-1945. Data pertaining to the nominees serve as a sort of reference point that allows us to better understand how winners are selected and how receiving the award affects them. Though we at first thought that centrality measures would provide a better indicator than citations, we observe that in fact both distributions are nearly identical. Interestingly, the peak in the nominees' ranks occurs slightly before the year of their nomination (4 years for physics, 1 for chemistry), while for the laureates, it occurs (on average and, in the case of chemistry, within the margin of error) the same year as their award. This is a surprising result, which remains valid (although much less pronounced, as is discussed below) in later periods. It is also somewhat counter-intuitive, given that the winners are selected many years after their discovery. Only four times were physics Nobel laureates awarded prizes within a year of their discovery, and the average lag time is around 12 years [Karazuai &Komkausaité, 2004]. In other words, this distribution suggests that while the impact of their experiment or theory might have been most important several years earlier, the Prize is awarded when their *accumulated symbolic capital* is highest [Bourdieu, 1975]. Furthermore, the shape of the curve corresponds closely to a normal distribution with an elongated time tail. This deviation for $t>0$ from the Gaussian can be interpreted as the "Halo Effect", not present in the case of nominees (whose names are not made public by the Nobel committee). This effect essentially reflects the law of cumulative advantage or the "Matthew effect" identified by the sociologist Robert K. Merton [1973]. Being recognized via a Nobel prize gives status to the scientist in the eye of his peers, who in turn accord more credit to him/her, which then translates into more citations. However, it is interesting to note that, on average, there is no evidence of this phenomenon generating an *increased* ranking in the years following the attribution of the prize, but simply a slower decline, shown in the asymmetry of the distribution.

Another important characteristic of these distributions is that the rate at which the ranking (in terms of citations or centrality) of scientists increases before the year of the Prize (i.e. the slope of Figures 1 and 2 at $t<0$), is much greater for winners than nominees. Thus, the former can be characterized as "rising stars" of the scientific community. It is also interesting to note that before about twelve years from getting their prize, the average ranking of winners is *lower* than that of the nominees, since winners cannot be nominated again once they have won. Thus, if two important discoveries (according to other scientists) are nominated for the Prize ten years thereafter, then the scientist who doesn't win will presumably be nominated again (if the validity of his work stands), so his $t \ll 0$ ranking will generally be higher. Take, for instance, the 1925 Nobel Prize in Physics awarded to James Franck and Gustav Hertz, for the well-known "Frank-Hertz experiments"[*], performed over 10 years earlier. That year, among the 16 nominees (who never went on to win the Nobel prize), we find figures such as Arnold Sommerfeld, Friedrich Paschen, Paul Langevin and Arthur Schuster, all of whom had been extremely well-known for at least twenty years and had already been nominated on several occasions. In essence, the fact that nominees are generally repeated year after year – while winners are selected once and usually at their "peak" – ensures that they have a relatively high standing (bibliometric ranking) over a longer period of time.

Most of the characteristics distinguishing nominees from winners described above are common to physics and chemistry, and this suggests that, at least for the first half of the $20^{th}$ century, these disciplines had a similar internal scientific dynamic for which citations and centrality offer a useful measure. As we mentioned at the beginning of this paper, the dynamics leading to the Prize in physiology and medicine seems more complex and less endogenous; citations and centrality measures are even less useful as predictive indices.

---

[*] These experiments demonstrated the electron's energy levels and thus supported Bohr's atomic model. Bohr himself had won the Nobel prize for his model only three years earlier, in 1922.

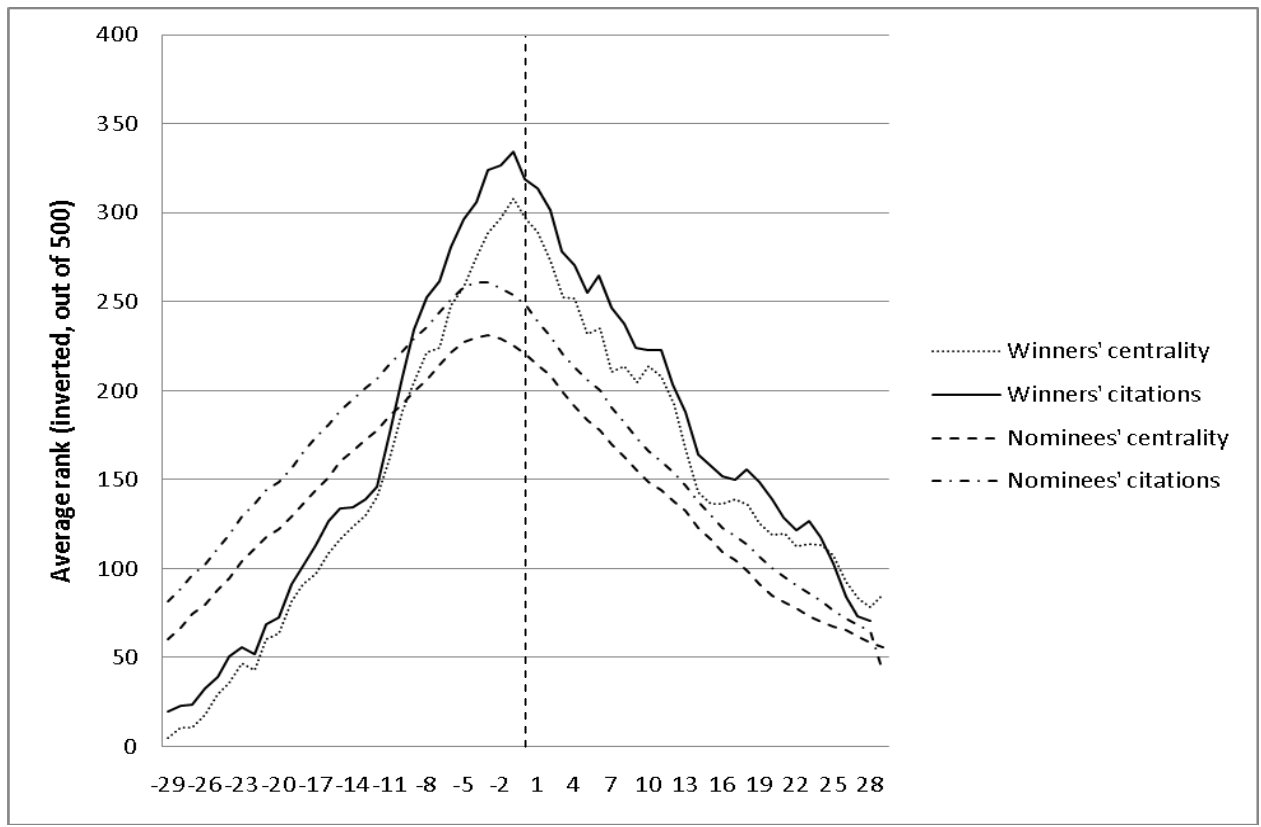

**Figure 1 : Average physics centrality and citation rankings for winners and nominees, 1901-1945. The vertical dashed line indicates the year "0", i.e. when someone is nominated or wins the award. Each data point therefore represents an average of all winners' rankings (between 1901 and 1945) at a given time interval from the Prize year. One can get an idea of the actual distribution of all rankings around the year the Prize is awarded, for instance, from Figs. 5 and 6.**

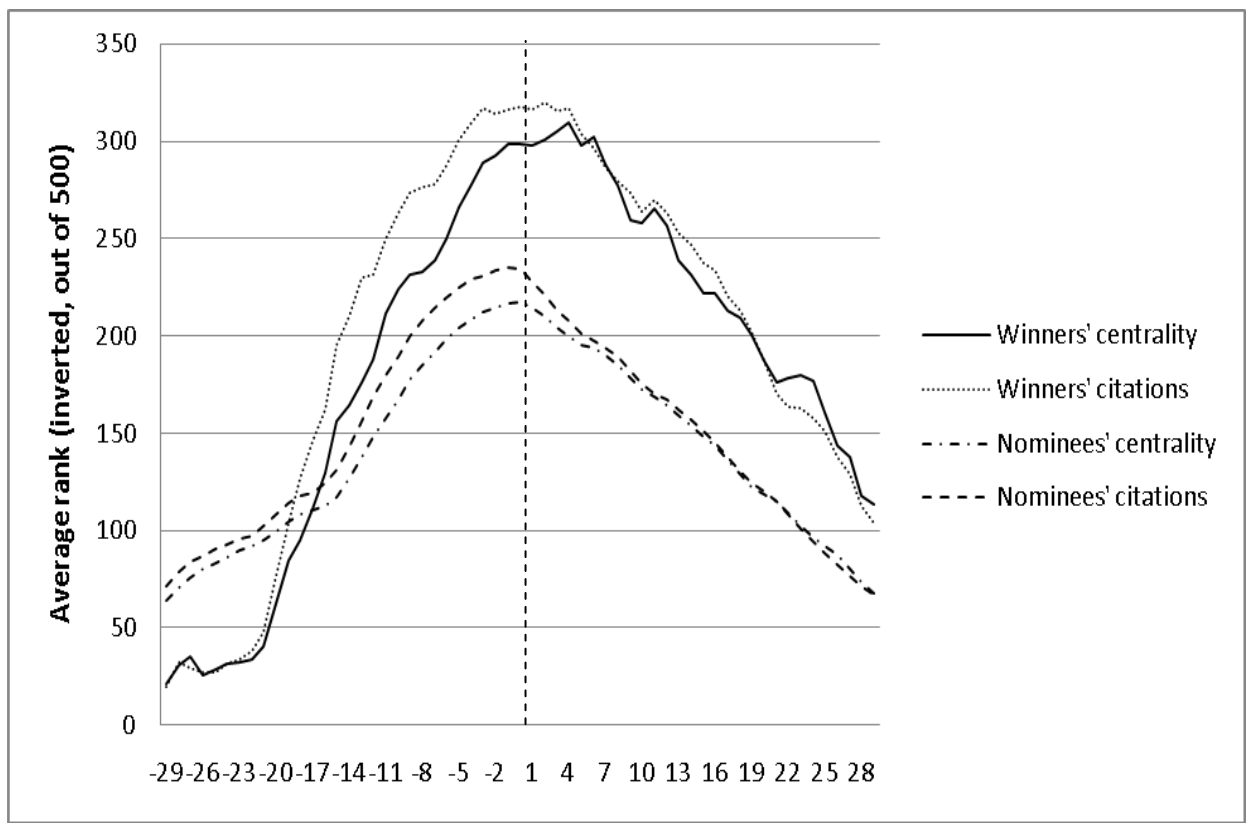

**Figure 2 :** Average chemistry centrality and citation rankings for winners and nominees, 1901-1945, following the same approach described in Figure 1.

**The Nobel Prize and the postwar growth in science: 1946-2007**

In order to inquire into the possible change over time in the dynamic of the scientific fields of physics and chemistry, we have divided the analysis of the distribution of citation rankings[*] into three periods (1901-1945; 1946-1970; 1971-2007). In this way, we can compare (Figures 3 and 4) the ranks of Nobel laureates in chemistry and physics over time. The first striking result is the progressive flattening of the distribution over the three periods. The rapid ascension of the winners as we approach the year of the Prize is replaced after 1970 by a nearly uniform distribution. The second period already suggests that the concentration of activities around a core of potential winners observed during the first half of the 20[th] century has rapidly changed to a situation where the community is fragmented into many small specialties. Nobelists thus become less distinguishable from the majority of top-level scientists. Although the lag between the publication of results and their ensuing Nobel Prize has been steadily increasing over the past 50 years [Karazuai & Komkausaité, 2004], there is no clear evidence of an important peak in citation or centrality rankings occurring *before* the Prize is awarded.

---

[*] The results using centrality rankings being similar, we omit them from the graphs.

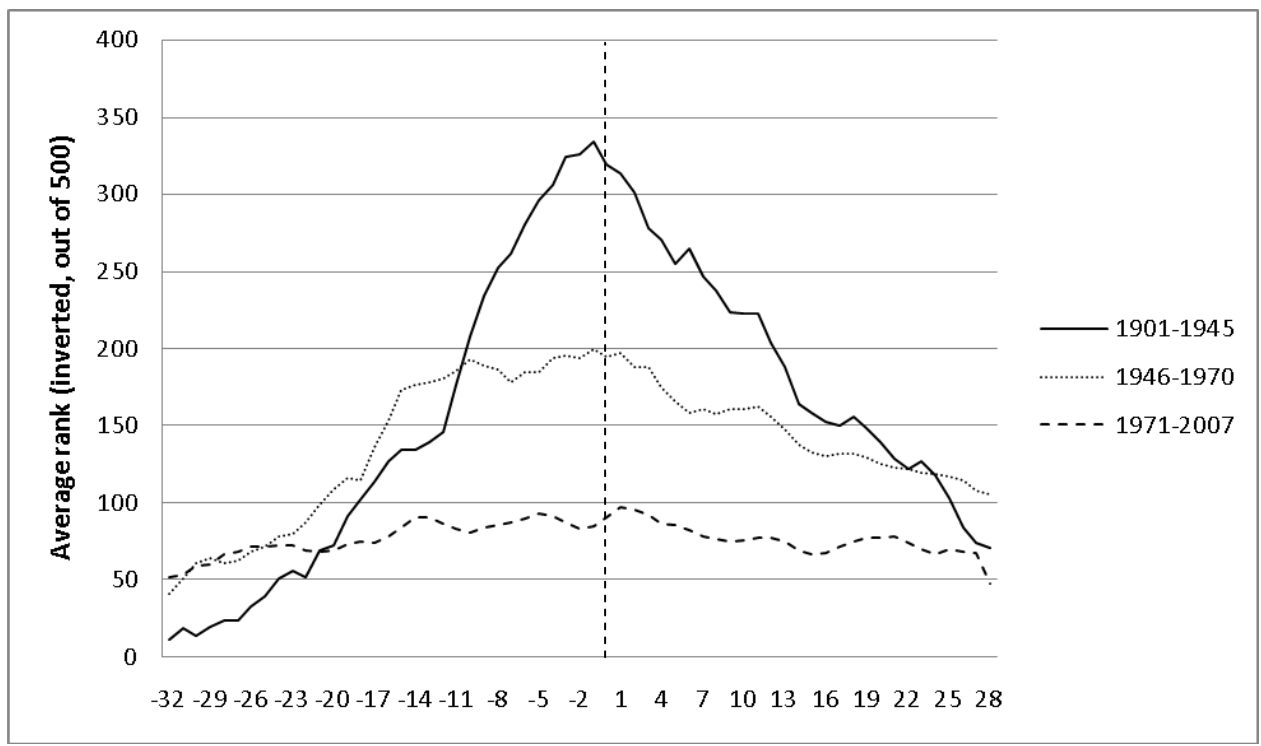

**Figure 3 :** Physics prizewinners' citation rankings, averaged over all years for three different periods. Once again, the vertical dashed line represents the year "0", when the Prize is won.

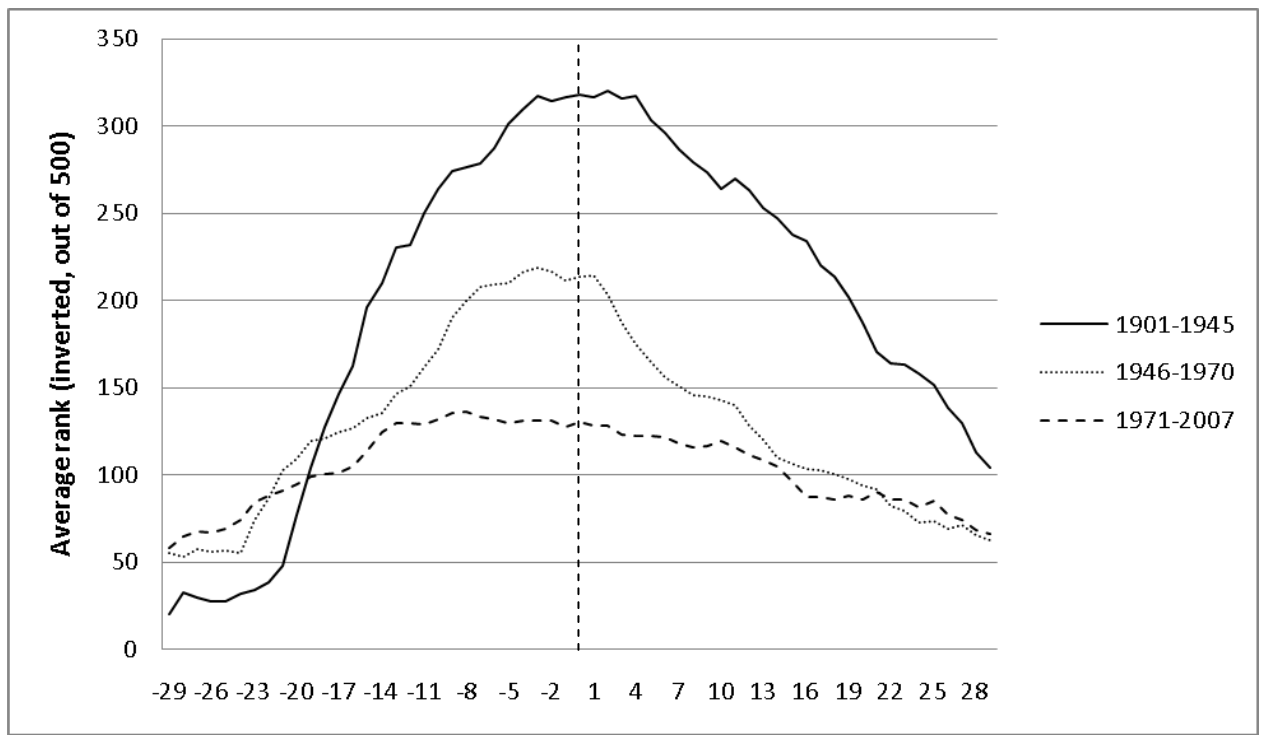

**Figure 4 :** Chemistry prizewinners' citation rankings, averaged over all years for three different periods. Once again, the vertical dashed line represents the year "0", when the Prize is won.

The most significant conclusion to draw from these distributions is that the predictive power of bibliometric measures over Nobel prizes has decreased over time and has now become greatly limited. Moreover, it is unlikely that improved techniques or approaches could remedy this situation since it stems, in the first place, from the explosion in the number of authors and the consequent fragmentation of disciplines into many relatively autonomous specialties: the Nobel committee can no longer easily isolate the most influential three physicists or chemists among a host of potential candidates. Whereas during the 1900-1945 period, a fairly large proportion of winners could be found in the top twenty or fifty most cited authors, the proportion of those being ranked above 500 become dominant after the 1970s *thus making the game of prediction almost futile*. The average rankings in themselves, however, do not provide complete information about the selection of prizewinners; as the peaks in Fig. 5 show, the *size* of the disciplines is a dominant, but not sufficient explanation. A finer analysis of our data is necessary in order to understand some of the other mechanisms at play in defining the post-war Nobelist's profile.

We therefore calculate the citation (and centrality) rankings during the three years preceding each prize, and analyze the distribution of the winners' ranks during each decade between 1901 and 2000. As shown in Figure 5 and 6, the odds of predicting the winner from the pool of candidates indeed becomes very low as time goes on. This is consistent with the rapid growth in the number of active scientists over the period as the probability of choosing a winner among the population is roughly inversely proportional to the number of scientists.

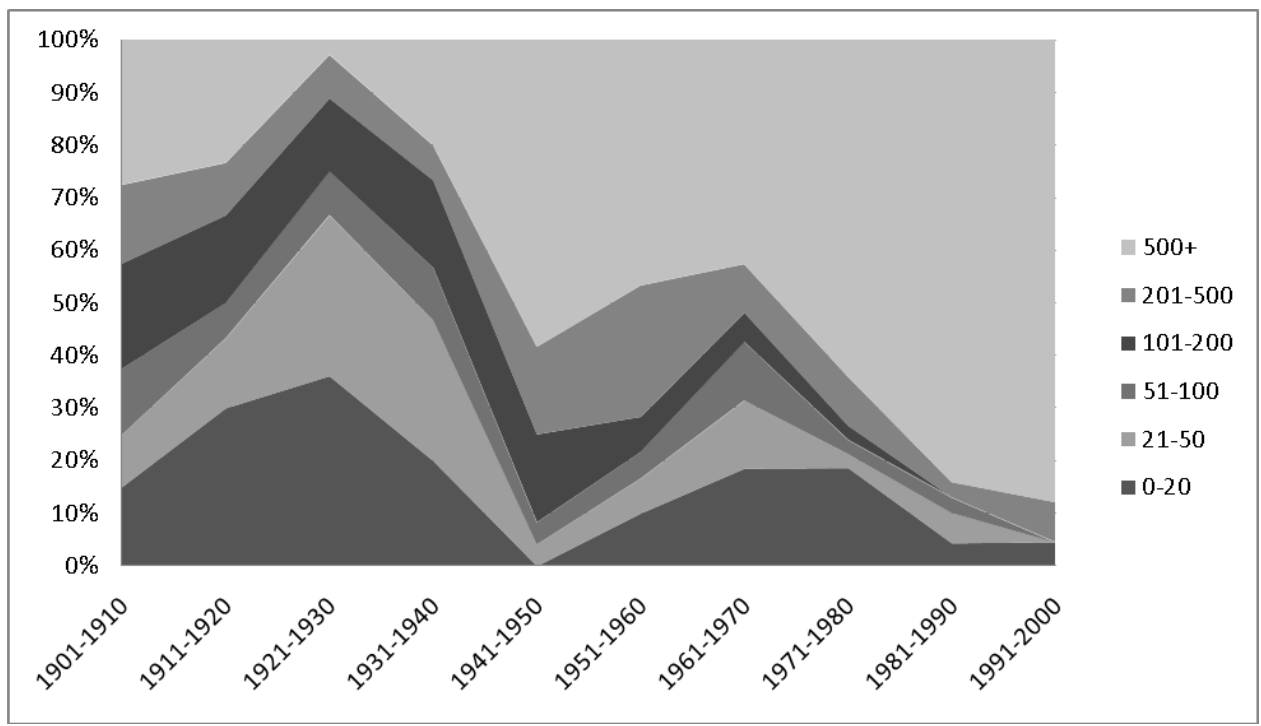

**Figure 5 : Distribution of citation rankings of physics prizewinners (by decade) compiled in the 3 years leading up to their being awarded the Nobel Prize. The six classes of rankings are chosen in order to compensate for the skewed distribution. Note the overall drop in highly-ranked scientists being awarded the Prize in later decades.**

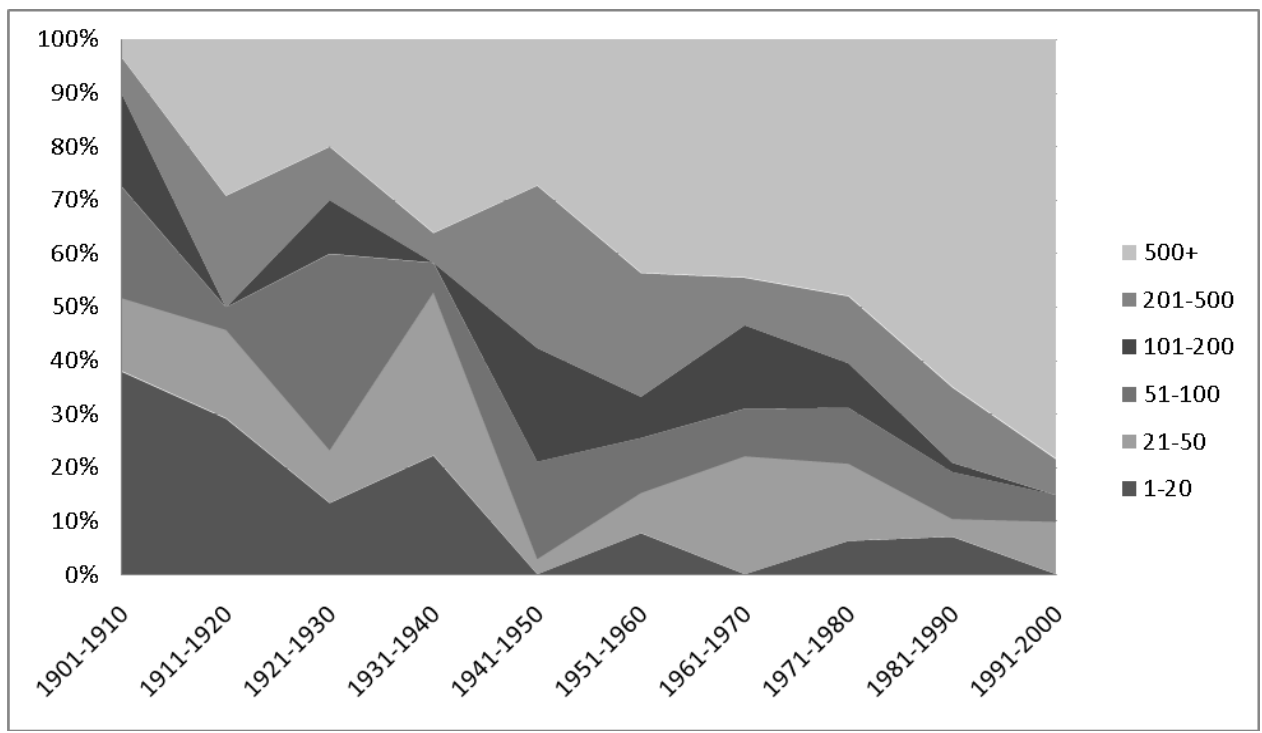

**Figure 6 : Distribution of citation rankings of chemistry prizewinners (by decade) compiled in the 3 years leading up to their being awarded the Nobel Prize. Note that while the same overall trend of increasingly lower-ranked prizewinners can be observed in this case (just as in physics), the "local" variations in the profiles of Nobelists in two disciplines are distinct.**

Let us examine more closely the case of physics in order to understand the laureates' rankings. First, we note that the highest ranks (especially the top 20) are those which vary most over the entire century. In other words, we will generally have a similar number of Nobelists ranked between, say, 200 and 500, but whether or not the "top" physicists are Nobelists is virtually impossible to predict.

In physics, the important fluctuations seen in Figure 5 can be explained by preferences for certain areas of the discipline. Table 1 breaks down the Nobel prizes between 1940 and 2000 according to the field (as defined by the PACS numbers [Karazuai & Komkausaité, 2004]) and a distinction between experimental and theoretical work. Once again, we use the rankings calculated over the three years before the Prize is awarded. Within the entire discipline, there is a clear hierarchy, at least in terms of how (and how often) work is cited: high-energy physics, density functional theory and semiconductors, for instance, occupy a relatively central position in physics, while astrophysics, optics, and the thermal and mechanical properties of condensed matter find themselves at the periphery. Similarly, theory is more central than experiment and usually ranks much higher in centrality as well as in citations. Garfield has also alluded to the presence of Nobelists from "smaller specialties": such recipients could have high rankings *within their specialties*, but not physics as a whole [Garfield & Welljams-Dorof, 1992].

The Nobel Prize, however, seems less dependent on this hierarchy of fields; there appears to be a will to distribute the prizes in a more "equitable" manner between specialties. The same phenomenon can be observed when we compare theoretical and experimental physics. The notion of "discovery" explains part of the lack of prizes (relative to their impact on the discipline) handed out to theorists: in almost all cases, the nobelists' theories must have already been rigorously confirmed by experiment before obtaining the Prize. On several occasions, the Prize has been jointly awarded to a theorist and an experimentalist, but the rank of the former is invariably higher than that of the latter. Two decades – the 1920s and 1960s – in which Nobel prizes in physics seem to include relatively high rankings laureates, reveal moments in time when the discipline is relatively compact and a few specialties are predominant thus making choices seem more obvious (see Figure 5 and Table 1). The rapid development of quantum mechanics in the 1920s is not only obvious to the historians of science today, but also to physicists and the Nobel selection committee at the time. Also, periods corresponding to the establishment of new central theoretical paradigms generate more theorists than usual among the Prize winners. The emergence of QED (quantum electrodynamics) just after the Second World War, and electroweak theories and high-energy physics (under the PACS classification of "elementary particles and fields") in the 1960s, explains the relatively high ranking of Nobel laureates during that period, as does the relatively high number of prizes awarded to theorists. The clear dominance of prizes given for experimental results since the 1980s seems to reflect the absence of major paradigm shifts in physics since the end of the 1960s. A similar type of analysis could probably explain the fluctuations in the chemistry prizewinners' curve, although each discipline has a different culture and internal dynamics.

In order to understand both the hierarchy of specialties and the fragmentation of the discipline, it is instructive to look at some specific, yet representative, examples. Take, for instance, the prizes awarded in 2001 and 2002, where the recipients were ranked extremely low in bibliometric terms *over the entire field*. In 2001, the award went to Ketterle, Cornell and Wieman for their (primarily experimental) work on Bose-Einstein Condensates only six years earlier. This is not to say that the their experiments went unnoticed or were seen as insignificant by other physicists, only that the work firmly established itself first within a niche of condensed matter experimentalists working to create Bose-Einstein condensates. In 2002, we find a similar case of primarily experimental astrophysics work[*] being rewarded without ever having been – according to the physics community as a whole – *central* to the discipline. In both cases, the selection of laureates was not done with respect to the discipline as a whole, *but rather with respect to the impact within a given specialty*.

| Classification | Average citation rank (inverted, out of 500) | Average centrality rank (inverted, out of 500) | 1941-1950 | 1951-1960 | 1961-1970 | 1971-1980 | 1981-1990 | 1991-2000 | Total |
|---|---|---|---|---|---|---|---|---|---|
| Condensed matter: structural, mechanical and thermal properties | 61 | 58 | 0 | 0 | 1 | 0 | 0 | 6 | 7 |
| Astrophysics and astronomy | 73 | 70 | 1 | 1 | 0 | 4 | 2 | 2 | 10 |
| General | 76 | 89 | 1 | 2 | 0 | 0 | 0 | 4 | 7 |

---

[*] The prize was shared between Davis and Koshiba for the detection of cosmic neutrinos, and Giacconi for work leading to the detection of cosmic X-ray sources.

| | | | | | | | | |
|---|---|---|---|---|---|---|---|---|
| Electromagnetism, optics, heat transfer, classical mechanics and fluid dynamics | 79 | 75 | 0 | 3 | 4 | 2 | 6 | 1 | 16 |
| Atomic and molecular physics | 141 | 133 | 0 | 2 | 2 | 3 | 4 | 3 | 14 |
| Physics of elementary particles and fields | 158 | 147 | 2 | 7 | 7 | 7 | 8 | 4 | 34 |
| Nuclear particles and fields | 196 | 183 | 4 | 2 | 3 | 0 | 0 | 0 | 10 |
| Condensed matter: electrical, magnetic and optical properties | 201 | 176 | 0 | 3 | 1 | 9 | 3 | 2 | 18 |
| TOTAL | - | - | 8 | 20 | 18 | 25 | 23 | 22 | 116 |
| **Experimental** | **61** | **54** | **5** | **13** | **3** | **8** | **20** | **18** | **67** |
| **Theoretical** | **234** | **220** | **3** | **7** | **15** | **17** | **3** | **4** | **49** |

**Table 1 : Sixty years of Nobel prizes in physics, broken down according to specialty and the (primarily) theoretical or experimental nature of the scientist's work. Note that the Prize can be jointly awarded to an experimentalist and theoretician for the same discovery.**

**Conclusion**

Our comprehensive bibliometric analysis of the profile of Nobel prizewinners in chemistry and physics from 1901 to 2007, showed that the changing dynamic of science due to its rapid increase in size since at least the 1960s, had the effect of diluting potential winners in a massive group of central scientists, from which it has become nearly impossible to pick up only three winners using bibliometric tools. We have shown that using rankings based on citations and obtained from Freeman's degree centrality in the cocitation network as indicators, give the same results. This is important not only in terms of understanding the limits of bibliometrics, but also for gaining insight into the social contexts and hierarchies that exist within scientific disciplines. We have also found that the distribution of ranking peaks at around the time the Prize is awarded and that a Halo Effect is consistently observable in the years following the attribution of the Prize. However, it is interesting to note that the "popularity" brought about by the Prize appears to be less important than the importance attributed to scientists *before* they become Nobel laureates. Furthermore, we are able to benchmark the laureates against other nominees in order to understand the "bibliometric" difference between the two.

Changes in the size and organization of the two fields result in a rapid decline of predictive power of bibliometric data over the century as the winners are distributed over a larger spectrum of rankings than at the beginning of the 20th century. This can be explained not only by the growing size and fragmentation of the two disciplines, but also, at least in the case of physics, by an implicit hierarchy in the most legitimate topics within the discipline, which is reflected more in bibliometrics than in the selection of the laureates. Further research could show whether or not normalizing the rankings within given specialties increases the odds of picking winners. Even then, the large number of such specialties and the existing limit of three winners for the discipline mean that the predictive power of bibliometrics will inevitably stay very low and that the Nobel Committee must take into account the politics of the discipline when choosing which specialty will be crowned in a given year.